\documentstyle[12pt]{article}

\begin{document}
\title{\large \bf Kinetic Roughening in Surfaces of Crystals Growing on
Disordered Substrates   }
\author{\normalsize Yan-Chr Tsai and Yonathan Shapir           \\
        \normalsize Department of Physics and Astronomy \\
        \normalsize University of Rochester \\
        \normalsize Rochester, N.Y. 14627,}
\maketitle
\normalsize
\begin{center} {\bf Abstract} \end{center}
\small

Substrate disorder effects on the scaling properties of growing crystalline
surfaces in solidification or epitaxial deposition processes are investigated.
Within the harmonic approach there is a phase transition into a
 low-temperature (low-noise) superrough phase with a continuously
 varying dynamic exponent
 $z>2$ and a non-linear response. In the presence of the KPZ
 nonlinearity the disorder causes the lattice effects to decay on large scales
with an intermediate crossover behavior. The mobility of the rough surface
  has a complex dependence on the temperature and the
 other physical parameters.

\newpage
\normalsize

 Much progress has been achieved recently in the understanding of kinetic
 roughening in nonequilibrium surface growth.$^{1-13}$ In particular spatial
and
 temporal scaling properties have been predicted for surfaces
 of crystals growing
   either by solidification  or from epitaxial deposition.$^{1,2}$ The simplest
analytical
 models neglect the discrete structure of the crystal and the scaling
 properties are
 determined from the university classes of the continuous kinetic equations
 which govern the growth process.$^{3-5}$

The lattice structure$^{6-13}$ has, however, crucial effects on the behavior
with
 a kinetic phase transition mirroring the equilibrium roughening
transition$^{14}$
 (ERT).
The transition is governed by the noise, i.e. thermal and  from the inherent
 stochasticity, of
 the growth process.$^{5-7,13}$ In presence of a strong noise the surface is
rough (in analogy with
 the rough equilibrium phase for $T>T_{r}$, where $T_{r}$  is the ERT
 temperature). In this phase the
 lattice effects are unimportant and the dynamic properties are the same as
 in the respective continuum models.
 In the low-noise phase the surface is smooth on an  intermediate scale (which
becomes larger the smaller  the
 average growth rate is) with
 drastically different dynamic properties.$^{6,7}$

 In view of the crucial importance of the discrete structure at the low-noise
regime, the following question should be addressed: What if  the substrate on
which
the solid grows  is not perfectly smooth? While all previous studies have
 assumed a perfectly flat substrate, in many potential realizations that will
 not be the
case. The effects of quenched disorder in the substrate on the kinetic scaling
properties of the growing crystalline surface are the subject of the present
letter. As detailed below we find
 that in presence of  substrate disorder   the low-noise (or low
 temperature) regime  has  dramatically different  kinetic properties.$^{15}$

The scaling properties  are manifested in the
 height-height correlation function:
\begin{equation}
C(L,\tau )=\langle \Delta h^2(L,\tau )\rangle =\langle [h(\vec {x}+\vec
{L},t+\tau )-h(\vec {x}
,t)]^2\rangle \sim L^{2\alpha}f(\tau /L^z)
,\label{eq:Co }
\end{equation}
$h(\vec{x},t)$ is the height of the surface at point $\vec {x}$ and time $t$.
The roughness
exponent $\alpha$ characterizes the long-time limit of the self-affine fractal
structure as represented
by the "surface width" $w(L,\tau )={\langle \Delta h^2 (L,\tau )\rangle}
^{1/2}\approx L^{\alpha}$.    At early stage of the growth ($\tau \ll L^{z}$)
$w\sim \tau ^{\beta}$   where
$\beta=\alpha/z$ and $z$ is the dynamic  exponent.
Theoretical studies of kinetic surfaces have been conducted within two
 approaches:  the "harmonic" approach$^{3,6,7}$ and the "non-linear"
 approach.$^{4,13}$ While
the latter takes into account the lateral growth of an oblique surface, the
 former applies to  discrete systems for which the nonlinear effects are
 negligible.$^{16-18}$ Our subsequent analysis follows this tradition.

I.\underline{Harmonic approach.}

The kinetic equation for the time evolution of the height  within this
 framework is:
\begin{equation}
\mu ^{-1}\frac {\partial h(\vec{x},t)}{\partial t}=F+\nu \nabla ^2 h(\vec{x},t)
 -\gamma_0 y
\sin [\gamma_0( h(\vec {x},t)+d(\vec {x}))] +\zeta (\vec {x},t)
,\label{eq:dis}
\end{equation}
where $\mu$ is a microscopic kinetic coefficient; $F$ is the driving force
 proportional
to the difference in chemical potential for solidification or determined by the
rate of deposition for epitaxial growth; $\nu$ is the surface tension; $y$ is
 the coefficient of the leading term due to the discreteness (higher harmonics
 are less
 relevant$^{6}$) and $\gamma_0 =\frac{2\pi}{b}$  where $b$ is the vertical
lattice spacing;
$\zeta(\vec{x},t)$ is the noise term with $\langle \zeta (\vec{x},t) \zeta (
\vec{x'},t') \rangle =2D \delta (\vec{x}-\vec{x'})\delta (t-t')$.
Note that this equation of motion can be derived from an Hamiltonian
$\mu^{-1}\frac{
\partial h}{\partial t}=- \frac{\partial {\cal H}}{\partial h}+\zeta$, detailed
 balance is obeyed, and at (or slightly off) equilibrium the temperature of
 system is $T=D\mu$.  Eq.(2) controls the
 lattice growth upon any generic rough surface, whatever the origin of the
imperfections is. We consider the finite quenched fluctuations  $d(\vec{x})$ of
the
 substrate height (for a schematic description see Fig. 1.) to have
 only short range correlations and to be at least of the order of the lattice
 spacing of the bulk solid $b$ (For $|d(\vec{x})|<<b$ the disorder is
 irrelevant). Defining $\theta(\vec{x})=\gamma_0 d(\vec{x})$, the phase-like
variables $\theta(\vec{x})$ obey
$\langle e^{i\theta(\vec{x})} e^{-i\theta(\vec{x'})}\rangle=a^2 \delta^2 (
\vec{x}-\vec{x'})$, where $a$ is the lattice spacing in the horizontal planes.

 In absence of disorder $ (d(\vec{x})=0)$, lattice effects have been studied
 within the harmonic
framework by Chui and Weeks$^{6}$ (CW) and later by Nozieres and Gallet$^{7}$
(NG). Their
important findings were: For $D\mu=T>T_{r}$, the kinetic (as the static)
 behavior is unaffected by the lattice potential $C(L,\tau)\sim (\ln L)
f(\tau/L^2)$(corresponding to  $\alpha =\frac{3-d}{2}=0$, $z=2$) and the
macroscopic
 mobility $\mu_{M}=\lim _{F\rightarrow 0} \frac{v}{F}$  ( $v=<\frac{
\partial h}{\partial t}>$- the average growth rate) is finite. For $T<T_{r}$,
 in the smooth phase, the
 surface tends to be pinned at the periodic
 minima, and $\mu_M=0 $ (with a finite jump at $T= T_r$). For
 finite $F$, therefore,  the growth is "activated"
 with nucleation$^{19,20}$ of higher "islands".

To apply the  renormalization group (RG) approach to Eq.(\ref{eq:dis})  with
the
 substrate disorder, we use the Martin-Siggia-Rose (MSR)
formalism.$^{21}$ In addition to $h(\vec{x},t)$, an auxiliary field
$\tilde{h}(\vec{x},t)$ is introduced as well as their
 conjugate "sources" $J(\vec{x},t)$ and $\tilde{J}(\vec {x},t)$.
The thermal noise and
  the quenched disorder are averaged upon to yield
 the following averaged generating functional:
\begin{eqnarray}
Z[J,\tilde {J}]&=&\int {\cal D}\tilde {h}{\cal D}h \exp\{\int\int d^2\vec{x} dt
 [Jh+\tilde{J}\tilde{h}+ D\mu ^2 \tilde {h}^2-\tilde
{h}(\frac {\partial h}{\partial t}-\mu \nu \nabla ^2 h)] \nonumber \\
               & &+\frac {\mu ^2\gamma ^2 g}{2a^2}\int \int \int d^2\vec{x} dt
dt'\tilde {h}(\vec {x},t)\tilde
{h}(\vec {x},t')\cos[\gamma(h(\vec {x},t)-h(\vec {x},t'))]\}\nonumber \\
                & & +\frac{1}{2} \mu^2 \bar{\nu}\int\int\int d^2\vec{x}dt dt'
\vec{\nabla}\tilde{h}(\vec{x},t) \vec{\nabla}
\tilde{h}(\vec{x},t'),
\label{eq:msr }
\end{eqnarray}
with $g=y^2a^4 $. The last term has been included since it is generated under
renormalization.

 Our RG analysis of this effective field theory follows closely that
of Goldschmidt and Schaub for the XY model with random anisotropies$^{22}$
(details of the calculation will be published elsewhere$^{23}$). Under
rescaling
  $x\rightarrow e^l x$ and $t\rightarrow e^{lz} t$ ( $l=\ln b$ where $b$  is
 the rescaling factor) the RG analysis yields, to lowest nontrivial order in
$g$, the following
 recursion relations:
\begin{eqnarray*}
\frac {d\nu}{dl}&=& 0 \qquad\qquad\qquad\qquad\qquad\qquad\qquad\qquad\qquad
\mbox{(4a)}  \\
\frac{d\bar{\nu}}{dl}&=& \frac {\pi\gamma^2}{4\nu (D\mu)^3} g^2 \qquad
\qquad\qquad\qquad \qquad
\qquad\;\;\;\;\;\;
\mbox{(4b)}\\
\frac {dF}{dl}&=&2F
\qquad\qquad\qquad\qquad\qquad\qquad\qquad\qquad\;\;\;\;\mbox{(4c)}  \\
\frac {dD}{dl}&=&(2-z+\frac {g\gamma^2  \sqrt{c}}{D\mu \nu })D \qquad\qquad
\qquad\qquad\qquad\:\mbox{(4d)}\\
\frac {d\mu }{dl}&=& (z-2- \frac {g\gamma^2  \sqrt{c}}{D\mu \nu })\mu
\qquad\qquad\qquad\qquad\qquad
\;\:\mbox{(4e)}\\
\frac {dg}{dl}&=& (2-\frac {\gamma^2 D\mu }{2\pi\nu })g-\frac {2\pi }{(D\mu
)^2}g^2.   \qquad\qquad
\qquad\;\;\;\;\;\mbox{(4f)}
\end{eqnarray*}
\begin{equation}
.\label{eq:re}
\end{equation}
The last equation provides the other parameter of the expansion $\delta=\frac{
\gamma^2 D \mu}{4 \pi \nu}-1$ (the deviation from the critical point) and these
 equations are of the first order in $\delta$. $\gamma$ remains at its bare
value $\gamma=\gamma_0$ and $\nu $ at $\nu_0$ (we may rescale $h$ such that
$\nu_0=1$).
We first look at the static properties by defining $T=D\mu$. It obeys
$\frac{dT}
{dl}=0$  and the last
 equation can be written as: $\frac{dg}{dl}=2(1-\frac{T}{T_{sr}})g-\frac{2 \pi}
{T^2} g^2$ with $T_{sr}=\frac{\nu b^2}{\pi}$.
We recognize the static equation of Cardy and Ostlund for the random-anisotropy
XY model.$^{24}$ Toner and DiVincenzo have analyzed these equations in the
context of
 equilibrium crystal surfaces with bulk disorder in a limit where their bulk
disorder is equivalent to the substrate disorder
considered here.$^{25}$ So their results directly apply: For  $T>T_{sr}$, $g
\rightarrow 0$, the disorder
is irrelevant,  and $w(L)\sim (\ln L)^{1/2}$. For $T<T_{sr}$, $g$ approaches a
 finite value $g^{\ast}\sim -\delta=1-\frac{T}{T_{sr}}$, disorder is relevant,
and the surface becomes superrough$^{25}$ $w(L)\sim \ln L$.

We now turn to study the kinetic properties. In the
 high-noise regime ($\delta >0$) we find $z=2$.  In the low-noise regime $z$
increases
 (the width spreading becomes slower)
 continuously as $z=2+4\sqrt{c}|\delta|$ with $c=\frac{1}{4}e^{2E}  \sim
0.7931$, where $ E$ is the Euler constant.
The mobility  $\mu_{M}$  far from the transition in the high $D\mu(=T )$ noisy
 phase  assumes
a finite value. Approaching $T_{sr}$ from above, however, $\mu_{M}$ vanishes
with
 $\delta$ continuously.
Integration of the recursion relation yields $\mu_{M}=\frac{\mu_{0}}{2\pi
g_0}|\delta|^{\eta}$
with $\eta=2\sqrt{c}\sim 1.58$. In the low-noise phase $\mu_{M}$ vanishes
  as $\frac{v}{F}\sim \frac{v_0}{F_0}e^{-(z^{\ast}-2)l}\sim
e^{-4\sqrt{c}|\delta|l}$. The behavior of $z$ and $\mu_M$ as function of $T$
is summarized in Fig. 2.

Physically the dynamic behavior in the superrough  phase may be understood
 as follows: the superrough phase is a phase in which the
 disorder barely
 dominates over the thermal fluctuations.$^{25}$ The increased roughness is the
result
 of the surface attempt to balance the tendency of the surface to adjust to the
 substrate (making $\vec{h}+\vec{d}$ an integer multiple of $b$) without paying
too much in elastic energy. Although the pinning is
 not uniform as with a smooth substrate the effect of locally preferred
locations
 is
 enough to slow the spreading of the "surface width "(as manifested by $z>2$)
 and to
prevent it from moving with a uniform  average velocity when an infinitesimal
driving force is
 $F$ applied.
 Naturally if a finite force is applied the surface will move, on the average,
at a constant velocity. This motion will wipe out the pinning effect
 (as it does to the periodic potential in absence of disorder). If the force
$F$
is small the behavior described here will hold up to a scale $L< L^{\ast}\sim
aF^{-1/2}$ and the effective mobility will be
 $\mu(l^{\ast}=\ln L^{\ast})\sim \mu_0 (L^{\ast})^{-4\sqrt{c}|\delta|}\sim
\mu_0 F^{2\sqrt{c}|\delta|}.$ It should be emphasized that this yields another
important finding namely the non-linear response to a small $F$ in
 presence of which the averaged velocity scales as $F^{2\sqrt{c}|\delta|+1}$.
This explicit behavior was derived based on scaling near the critical point but
is also consistent with
 activated dynamics over free-engery barriers$^{26}$ $E(L)$ given by $\epsilon
(T)\ln L
$ with $\epsilon (T)=4\sqrt{c}(T_{sr}-T)$.

II.\underline{The "non-linear" approach.}

Kardar, Parisi, and Zhang (KPZ) have pointed out that when the lateral growth
of
an oblique surface is accounted for the most relevant effect is the addition of
a
term of the form $\frac{\lambda}{2}(\nabla h)^2$ to the growth equation.$^{4}$
In
 absence of any lattice or
 disorder effects the KPZ equation is $\mu^{-1}\frac{\partial h}{\partial
t}=F+\nu
\nabla^2 h+\frac{\lambda}{2}(\nabla h)^2 +\zeta (\vec{x},t)$, and the
 exponent $\alpha$ and $z$ change$^{1}$ from $\alpha=0$ and $z=2$ for
$\lambda=0$, to
 $\alpha \sim 0.4$ and $z\sim 1.6$.

The question of whether a phase transition may occur in presence of
 nonlinearity
 and a lattice  with a perfectly flat substrate has been considered in a number
of recent simulation of
deposition (or growth) of discrete particles.$^{8-12}$
 E.g., the observed transition between logarithmic and power-law behavior of
$w^2(L)$
 by Amar
 and Family$^{8}$ has been attributed to an effective vanishing of the
nonlinear
term in their discrete model.$^{16-18}$
An extensive analytic study which includes both the lattice effects and the
 nonlinearity was performed by Hwa, Kardar and Paczuski$^{13}$ (HKP).
Studying the intermediate scale $L<L^{\ast}$, HKP identified two phases.  One
is
 the high temperature (strong noise) rough phase crossing over to  KPZ
scaling.$^{13}$ Approaching
the transition from this phase the mobility vanishes as $(\ln |T-T_c|)^{-\zeta
^{'}}$. They also argue
 that for  $T<T_{c}$ (lower noise) the surface is flat. This
 identification
requires some caution since the generation of a term $y_2 \cos(\frac{2\pi}{b}h)
$  from the contraction
of the terms $y_1 \sin (\frac{2\pi}{b}h)$  and $\frac{\lambda}{2}(\nabla h)^2$
  was not considered. Combining both terms into the form $|y|\sin(\frac{2\pi}
{b}h+\theta(l))$ (with $y^2=y_1^2 +y_2^2$ and $\theta (l)=tg^{-1} \frac{y_2}
{y_1}$)
, we find that for $T<T_c$  the flow is indeed toward $|y|\rightarrow \infty$
  but the phase shift angle
 is rotating  like  $\theta(l)=\omega l$ with  $l$, $\omega \sim \lambda/\nu$.
 Thus, this low temperature phase is not  characterized simply by an
  increase of the periodic potential. Higher order terms in the
recursion relations  of   $y_1$  and $y_2$ and that of $\nu$  will be required
to identify with more confidence
the nature of the low-temperature phase.

Turning  now to the effect of substrate disorder in presence of the KPZ
nonlinearity:
 The RG analysis becomes much more complex. It turns out that a systematic
expansion in the parameter  $\delta =\frac{D\mu\gamma^2}{4\pi}-1$  requires
 the consideration of diagrams containing
 up to three non-trivial loops.  Sophisticated techniques, based on dimensional
regularization,
were employed to extract their singular
 parts.$^{23}$
  The following recursion
relations for  $x\rightarrow x e^l$   $t\rightarrow t e^{2l}$  were derived
($\gamma =\gamma_0$)
\begin{eqnarray*}
\frac{d\nu}{dl}&=&0 \qquad\qquad\qquad\qquad\qquad\qquad\qquad\qquad\qquad
\;\;\;\;\:\mbox{(5a)} \\
\frac{d\bar{\nu}}{dl}&=& \frac {\pi\gamma^2}{4\nu (D\mu)^3} g^2 \qquad
\qquad\qquad\qquad \qquad
\qquad\;\;\;\;\;\;\;\;\;\:
\mbox{(5b)}\\
\frac{dF}{dl}&=& 2F+\pi\lambda \qquad\qquad\qquad\qquad\qquad\qquad\qquad\qquad
\mbox{(5c)} \\
\frac{dD}{dl}&=&(\frac{\lambda^2}{8\pi}D\mu+\frac{\gamma^2\sqrt{c}g}{D\mu})D
\qquad\qquad\qquad\qquad\qquad\;\;\;\: \mbox{(5d)} \\
\frac{d\mu}{dl}&=&(-\frac{\gamma^2 \sqrt{c}g}{D\mu})\mu
\qquad\qquad\qquad\qquad\qquad\qquad\qquad
\;\;\mbox{(5e)} \\
\frac{dg}{dl}&=&(2-\frac{D\mu \gamma^2}{2\pi}-\frac{\lambda^2 c'}{\gamma^2}
)g-\frac{2\pi}{(D\mu)^2} g^2 \qquad\qquad\qquad \mbox{(5f)}\\
\frac{d\lambda}{dl}&=&0,
\qquad\qquad\qquad\qquad\qquad\qquad\qquad\qquad\qquad\;\:\; \mbox{(5g)}
\end{eqnarray*}
\begin{equation}
.\label{eq:ere}
\end{equation}
where $\nu_0$ is set to be 1, and $c' \sim 180.08$.
Combining the equations for $D$  and $\mu$  together we find ($T=D\mu$):
\begin{equation}
\frac{dT(l)}{dl}=\frac{\lambda ^2}{2\gamma ^2} T(l)
,\label{eq:tem}
\end{equation}
\begin{equation}
\frac{dg(l)}{dl}=(2-\frac{T(l)}{2\pi}\gamma^2-\frac{c'\lambda^2}{\gamma^2})g(l)
-\frac{\gamma^4}{8\pi^2} g^2(l)
.\end{equation}
The most important observation is that the nonlinearity increases the effective
temperature $T(l)$ and,  as a result, the effective coupling $g(l)$  becomes
smaller. The only
 fixed point has $T\sim T_0 e^{\frac{\lambda ^2}{2\gamma^2}l}\rightarrow
\infty$
  and therefore $g\rightarrow 0$.
The equation for $g(l)$  can be integrated exactly:
\begin{equation}
\frac{1}{g(l)}=\frac{1}{g_0}e^{s(l)}-\frac{\gamma^4}{8
\pi^2}e^{s(l)}\int_{0}^{l}dx e^{-s(x)},
\label{eq:tgt}
\end{equation}
 where $s(x)=(\frac{\lambda^2 c'}{\gamma^2}-2)x+T_0\frac{\gamma^4}{\pi
\lambda^2}(e^{\frac{\lambda^2}{2\gamma^2}x}-1)$.
  The behavior on long scales will have $g\rightarrow 0$ and the effective KPZ
 coupling $K=\frac{\lambda^2 D \mu}{\nu^3}$ [note
 that to these orders in $\delta$, $g$  and $\lambda^2$, the flow of this
coupling  is unaffected by  $g$ ]
 will control the behavior.

However, since the scale associated with the increase of the KPZ coupling is
exponentially large $L_K =ae^{\frac{8\pi}{K_0}}$  it is likely to  be larger
than $L^{\ast}$ (the scale set by $F$). Therefore on scales smaller than
$L^{\ast}$ a rough surface will be observed
 but it will be in a crossover regime.
 If  $\lambda _0$ and/or $T_0$  are small (especially with $g_0$ large), $g(l)$
 will decay to zero quite slowly. This will be observable in the mobility which
has the scale dependence
$\mu(l)=\mu_0\exp\{-\frac{\gamma^4 \sqrt{c}}{4\pi}\int_0^{l} dl' g(l')\}$
with $g(l)$ given in Eq.(\ref{eq:tgt}).
 $\mu_M\sim\mu(l^{\ast}=\ln L^{\ast})$  will
be  drastically reduced by the disorder effects (as they decay on intermediate
scales) compared with its bare value $\mu_0$, the same will hold for the width
 $w(l)$ when compared with the $g=0$ case.

To summarize we have investigated how the scaling properties of growing
 crystalline surfaces are affected by disorder in the substrate.  For
$T<T_{sr}$ in the harmonic approach,  the surface is
 superrough with anomalous dynamics. The height-height correlations are
$C(L,\tau)\sim(\ln L)^2 f(L/\tau ^z)$
with $z=2+4\sqrt{c}(1-\frac{T}{T_{sr}})$.
 At the same time the  response becomes nonlinear: $v\sim F^{\zeta+1}$ with
 $\zeta=2\sqrt{c}(1-T/T_{sr})$.
 In presence of the KPZ non-linearity  the complex reduction in the  width and
the  mobility of the rough surface have been obtained.
These effects may be discernible in future precise measurements of
solidification and epitaxial deposition
processes.

\newpage
\underline{Acknowledgments}

We are thankful to T. Hwa, M. Kardar, G. Grinstein, B. Schmittmann
 and especially
 to D. Huse for most useful discussions. We  are also indebted to Y.
Goldschmidt for
bringing to our attention Ref. 22.
This work was partly supported by a grant from the Corporate Research
Laboratories of the
Eastman-Kodak Company in Rochester.

\newpage
\noindent{\large\bf Figure Captions}

\noindent{\bf Fig. 1.}
- A two dimensional cut (along a lattice plane perpendicular to the
 disordered substrate) of the three dimensional system.

\noindent{\bf Fig. 2.}
- The dependence of the linear response  macroscopic mobility $\mu _M$ (bold
line) and the
  dynamic exponent $z$ (dashed line) on temperature for the harmonic model. The
arrows indicate the
 appropriate scales on the vertical axis. ({\bf  R} - the rough phase
$T>T_{sr}$,
 {\bf SR} - the superrough phase $T<T_{sr}$)


\begin{thebibliography}{99}
\bibitem{26} Implications to the dynamics of charge-density waves and vortex
 glasses will be discussed in Ref (23).
\end{thebibliography}
\end{document}